\documentclass[pre,aps,onecolumn,superscriptaddress,showpacs,floatfix]{revtex4}
\usepackage{graphicx}
\usepackage{subfigure}
\usepackage{amsmath, amssymb}
\begin{document}
\DeclareGraphicsExtensions{.eps}
\title{The out of equilibrium behavior of  Casimir type fluctuation induced forces for 
free classical fields }

\author{David S. Dean}
\address{Laboratoire de Physique Th\'eorique, IRSAMC, Universit\'e de Toulouse, UPS and  CNRS, 118 Route de Narbonne, 31062 Toulouse Cedex 4, France}
\author{Ajay Gopinathan}
\address{School of Natural Sciences,
University of California,
Merced, CA 95344, U.S.A.}

\begin{abstract}
We present a general method to study the non-equilibrium behavior of  Casimir type fluctuation induced forces  for classical free scalar field theories. In particular we analyze the temporal evolution
of the force towards its equilibrium value when the field dynamics is given by a general class of over damped stochastic dynamics (including the model A and model B class).  The steady state force is
also analyzed for systems which have non-equilibrium steady states, for instance where they are driven
by colored noise. The key to the method is that out of equilibrium  force is  computed by specifying an energy of interaction between the field and the surfaces in the problem. In general we find that there
is a mapping of the dynamical problem onto a corresponding static one, and in the case where the 
latter can be solved the full dynamical behavior of the force can be extracted.  The method is used how to  compute the non-equilibrium Casimir force induced between two parallel plates by a fluctuating field, in the cases of Dirichlet, Neumann and mixed boundary conditions. Various other examples, such as the 
fluctuation induced force between inclusions in fluctuating media are discussed.
\end{abstract}

\maketitle
\section{Introduction}
The Casimir effect arises when the fluctuations of a quantum field is modified by
the presence of surfaces or objects placed in the field  \cite{krech1994,kar1999,mos1997,mil2001,bor2009}. 
A so called pseudo or critical Casimir  force can also arise for classical fields in the presence 
of thermal fluctuations, and although the physical origin of this pseudo-Casimir force is quite different to
the quantum Casimir effect the two forces have a similar origin from a mathematical stand point.
The terminology used to describe these types of Casimir forces varies considerably in the literature.
We will be interested in the classical, non-quantum (or zero Matsubara frequency) Casimir interaction
which is dominant at high temperatures and large separations between interacting surfaces.
The interaction arises in the context of a statistical field theory rather than a quantum field theory
and is often referred to as the thermal Casimir effect as in this regime the force is proportional
to the temperature. When the statistical field theory describes a critical point, then one can 
talk precisely of a  critical Casimir effect. However even away from a critical point a Casimir type 
interaction occurs, but this interaction is screened rather being long-range.
The interaction induced between surfaces or objects can be considered to be due to the 
imposition of boundary conditions on the field or due to an energy of interaction with the field. 
The simplest system where the thermal Casimir (or pseudo-Casimir)  effect arises is  the  free scalar field
theory, and long range Casimir forces arise when the field theory is massless, {\em i.e.} where the Hamiltonian of the system is given by:
\begin{equation}
H = {1\over 2} \int d{\bf x} \ \left[\nabla\phi({\bf x})\right]^2.\label{ham}
\end{equation}
The above field theory is purely classical and is, for instance, a simple model for the elastic energy of a surface if the field $\phi$ represents the height of the surface. If one specifies the boundary conditions for the field on two plates ({\it e.g.} Dirichlet, Neumann or Robin) the equilibrium thermal Casimir interaction between the plates can be computed from the free energy \cite{kar1999,mos1997}. The  fluctuating field $\phi$ also describes the order parameter for critical systems, such as binary liquids at the critical point \cite{fi1978}, and the occurrence of such a critical Casimir interaction has recently been confirmed experimentally \cite{he2008}. The effect of the Casimir force on the wetting and thinning of  $^4He$
and  $^4He- ^3He$ mixtures has also been extensively studied, theoretically \cite{mac2007,zan2007},
numerically \cite{vas2007,hu2007} and experimentally \cite{gar1999,gas2006}. A recent review of  results on the critical Casimir force can be found in \cite{gamrev}. Also the field $\phi$ can  represent the phase of the complex order parameter in a superfluid state, such as that occurring for $^4He$ \cite{zan2004}, giving an additional contribution to the Casmir-like forces 
in wetting films of $^4He$.  We note that in general  critical systems  are described by interacting field theories, the approach in this paper
only applies to free field theories and can only be applied to interacting field theories as a Gaussian
approximation. Free vectorial field theories describing liquid crystal systems also exhibit Casimir type interactions between the surfaces confining the system \cite{ad1991}. In what follows we will study classical field theories where the  dynamics of the fields are driven by thermal fluctuations or other stochastic noise, i.e. we will consider a range of dissipative dynamics for the system. For brevity we shall refer to the forces as Casimir forces rather than pseudo Casimir or critical Casimir forces, and at no point will we consider quantum effects.

When classical fields  are out of thermal equilibrium, one expects significant changes in the Casimir forces from those computed at equilibrium. One of the principal problems when analyzing the 
out of equilibrium Casimir effect, say for parallel plates, is to obtain an expression for the force between the two plates which is valid out of equilibrium. In previous studies, the stress tensor has been used to study both the dynamical behavior of the force \cite{ba2003,na2004,ga2006} and the force fluctuations in equilibrium \cite{ba2002}. However, only the  force in equilibrium can be strictly computed using the stress tensor (see the later discussion), therefore there is no general proof that the stress tensor can be used to compute forces out of equilibrium for general dissipative dynamics. Results using the stress tensor may, however, be reliable for situations close to equilibrium \cite{ga2006,ga2008}. For the quantum Casimir interaction the field dynamics is Lagrangian
and progress is being made on its  out of equilibrium behavior \cite{bor2009}. 
Another approach to compute out of equilibrium Casimir forces is to construct a model with a specified non equilibrium dynamics and to specify by hand the force at the wall. For example, in \cite{br2007,bu2008}, the dynamical field was related to a particle density and the local pressure on the wall is then given by the ideal gas form via kinetic reasoning. We note that in a number of papers where the stress tensor is used out-off equilibrium  or where  kinetic arguments are used to compute the force
in parallel plate geometries it has been found that the forces exerted on the plates are not equal and
opposite. In some of these works \cite{ba2002,bu2008} this fact is interpreted as a violation of the  Newton action reaction principle. Given these intriguing and interesting results it is therefore useful to develop an analysis where we can write down the instantaneous force on the plates or surfaces in terms of the unaveraged field variables without making any further assumptions to evaluate the force..    

To achieve this goal, in this paper  rather than imposing boundary conditions on the field we specify its energy of interaction with the surfaces in the system (for the static problem see \cite{de2009} for example). In this way we can write down the instantaneous force on the wall unambiguously as all
forces in the problem are generated by a potential. Recently a variant of this approach was also
applied to compute dynamical drag forces in fluctuating classical fields \cite{dem2010}
This paper presents the full details of the 
calculations presented in an earlier letter \cite{dego2009} and in addition clarifies how our formalism 
for computing the force is related to the stress tensor. We also discuss the out of equilibrium force  between plates with mixed boundary conditions for parallel plane geometries. By using the 
pairwise approximation we also  show how this method can  be applied to compute the out of equilibrium fluctuation induced force between small inclusions in fluctuating media.

\section{General Formalism}
\label{secGF}
\subsection{Energetic formulation of the boundary interaction}
We commence by considering the most general case of a free field theory where the Hamiltonian can be
written in terms of a general quadratic Hamiltonian \cite{note1}. 
\begin{equation}
H ={1\over 2} \int d{\bf x} d{\bf x}' \phi({\bf x})\Delta({\bf x}, {\bf x}', l)\phi({\bf x}')\label{eqH},
\end{equation}
where $\Delta$ is a self-adjoint operator {\em i.e.} $\Delta({\bf x},{\bf x}')=\Delta({\bf x}',{\bf x})$.
Here $l$ represents any suitable free parameter in the problem but for concreteness it could be
the position of a plate which interacts with the field. For instance one could chose we choose
\begin{equation}
\Delta({\bf x}, {\bf x}',l) = -\left[ \nabla\cdot \kappa(z,l)\nabla -\delta(z) c_1 - \delta(z-l) c_2\right]\delta({\bf x}-{\bf x'}),
\end{equation}
where when $c_1$ and $c_2$ are positive this corresponds to a free field theory where the fluctuations of the field $\phi$ are suppressed on two plates, one at $z=0$ and the other at $z=l$. The term $\kappa(z,l)$  is a spatially varying {\em elastic constant} for the field, for instance one could have 
one value within the two plates and another outside. The induced boundary conditions for this theory 
at each plate are of the Robin form:
\begin{equation}
\kappa(0^+,l){\partial \phi\over\partial z}\vert_{0^+} - \kappa(0^-,l){\partial \phi\over\partial z}\vert_{0^-} 
=c_1\phi(0)
\end{equation}
at $z=0$ and 
\begin{equation}
\kappa(l^+,l){\partial \phi\over\partial z}\vert_{l^+} - \kappa(l^-,l){\partial \phi\over\partial z}\vert_{l^-} 
=c_2\phi(l),
\end{equation}
at $z=l$ and where the superscripts $x^\pm$ indicate being infinitesimally to the  right and left of the point $x$. Clearly, when $\kappa$ is constant  and in the limit where $c\to \infty$ one
will obtain Dirichlet boundary conditions on the two plates. The instantaneous generalized force acting on the plate at $z=l$ is thus given by
\begin{equation}
F_l = -{\partial H\over \partial l} =-{1\over 2} 
\int d{\bf x} d{\bf x}' \phi({\bf x}){\partial \over \partial l} \Delta({\bf x}, {\bf x}', l)\phi({\bf x}').\label{eqforce}
\end{equation}
This is the strict definition as defined by the principle of virtual work, it is valid for any configuration
of the field $\phi$ and position of the interacting surface or object. 
Clearly simply that the potential energy of a physical system is described
by a potential $V({\bf x})$ then the instantaneous force in the direction $i$  for any configuration of 
the coordinates
$$
f_i = -{\partial V\over \partial x_i},
$$
we are thus simply applying this result to functional potential energies.
The equilibrium value of this force is can also be written in the familiar form
\begin{equation}
\langle F_l\rangle = T{\partial\over \partial l}\ln(Z(\Delta)) \label{eqfe}
\end{equation}
where 
\begin{equation}
Z(\Delta) = \int d[\phi] \exp(-\beta H[\Delta] )
\end{equation}
with $H$ as defined in Eq. (\ref{eqH}) and where $\beta = 1/T$ is the inverse temperature. 
Alternatively the force can be expressed using Eq. (\ref{eqforce}) to obtain
\begin{equation}
\langle F_l \rangle =-{T\over 2} 
\int d{\bf x} d{\bf x}' \ \left[{\partial \over \partial l} \Delta({\bf x}, {\bf x}', l)\right]
\Delta^{-1}({\bf x}, {\bf x}', l) \label{eqfc},
\end{equation}
where we have simply used the fact that the static correlation function is given by
\begin{equation}
\langle \phi({\bf x})\phi({\bf x}')\rangle = T\Delta^{-1}({\bf x},{\bf x}').
\end{equation}

Before using the above idea to compute the Casimir force out of equilibrium we will briefly show how
the energetic formulation can be used to derive the standard expression for the {\em average} value
of the force at {\em thermal equilibrium}. This is a useful exercise as it shows that our method recovers the stress tensor result in equilibrium and it is also a useful reminder as to when the use of the stress 
tensor is valid. The more traditional derivation in the context of quantum field theory can be found for example in \cite{wein}.

Consider the Hamiltonian of a scalar field theory $H$ with Hamiltonian density $\cal{H}$ such that
 \begin{equation}
 H = \int  d{\bf x}\ {\cal H}
 \end{equation}
 and where the Hamiltonian density consists of  a bulk part denoted by ${\cal H}_0$ and an interaction
 term with a surface $S$ denoted by ${\cal H}_S$:
 \begin{equation}
 {\cal H} = {\cal H}_0 + {\cal H}_S.
 \end{equation}
 Here we assume that $H_0$ has a standard quadratic kinetic term $[\nabla\phi]^2$ plus
 an interacting term depending only on $\phi$. 
 For instance if the surface is perpendicular to the direction $i$ and is at $x_i=l_i$ then we can write
 \begin{equation}
 {\cal H}_S = \delta(x_i-l_i) V(\phi).\label{asurface}
 \end{equation}
 The force on the surface in the direction $i$ is then given by
 \begin{equation}
 F_i = -{\partial H\over \partial l_i} = -\int d{\bf x}\  {\partial {\cal H}\over \partial l_i}=-\int_{V_S}d{\bf x} {\partial {\cal H}\over \partial l_i}, \label{aforce}
 \end{equation}  
where the volume $V_S$ in the last integral represents any volume (which can be infinitesimally small)
containing the surface, this is because of the localized nature of the surface interaction given in 
Eq. (\ref{asurface}). 

In order to make the connection between our energetic formalism and the stress tensor we compute the
derivative of the Hamiltonian density $\cal H$ in the direction $i$ which can be written as
\begin{equation}
\nabla_i {\cal H} = {\partial \cal{H}\over \partial \phi}\nabla_i \phi + {\partial {\cal H}\over \partial\nabla_j \phi}\nabla_j\nabla_i \phi -{\partial {\cal H}\over \partial l_i}    
\end{equation}
we can now use this is Eq. (\ref{aforce}) to give 
\begin{equation} 
F_i = \int_{V_S} d{\bf x}\ \left[ \nabla_i{\cal H} - {\partial \cal{H}\over \partial \phi}\nabla_i \phi - {\partial {\cal H}\over \partial\nabla_j \phi}\nabla_j\nabla_i \phi \right].
\end{equation}
Now integrating the last term by parts we find that
\begin{equation} 
F_i = \int_{V_S} d{\bf x}\ \left[ \nabla_i{\cal H} - \nabla_j\left({\partial {\cal H}\over \partial\nabla_j \phi}\nabla_i \phi\right)
-\left({\partial \cal{H}\over \partial \phi}-\nabla_j\left({\partial {\cal H}\over \partial\nabla_j \phi} \right)\right)\nabla_i \phi \right].
\end{equation}
This can now be written as
\begin{equation}
F_i = \int_{V_S} d{\bf x}\ \nabla_j T_{ij} -\int_{V_S}d{\bf x}\ \left({\partial \cal{H}\over \partial \phi}-\nabla_j\left({\partial {\cal H}\over \partial\nabla_j \phi} \right)\right)\nabla_i \phi 
\end{equation}
where 
\begin{equation}
T_{ij}=  \delta_{ij} \mathcal{H} - \nabla_j\phi{\partial\mathcal{H}\over \partial \nabla_i\phi},
\end{equation}
is the standard stress tensor. We see that in general there is an additional term in the 
general force given by the second integral above but which clearly must be zero in 
equilibrium. To see how this vanishes at equilibrium we note that the argument of the
second integral above is in fact the functional derivative of the Hamiltonian $H$, {\em i.e.}
\begin{equation}
{\partial \cal{H}\over \partial \phi}-\nabla_j\left({\partial {\cal H}\over \partial\nabla_j \phi} \right)
= {\delta H\over \delta\phi}
\end{equation}
Note that in equilibrium the classical field obeys $\delta H/\delta\phi=0$, however strictly speaking to show the vanishing of the second integral in equilibrium we use the Schwinger-Dyson equation
\begin{equation}
\int d[\phi]\ {\delta\over \delta\phi({\bf x})}\left(\phi({\bf y})\exp(-\beta H)\right) = 0,
\end{equation}
as the functional integral is an exact derivative.
This  can now  be rearranged to give 
\begin{equation} 
\langle \phi({\bf y}){\delta H\over \delta\phi({\bf x})}\rangle = {1\over \beta}\delta({\bf x}-{\bf y})
\end{equation}
where the angle brackets indicate averaging with respect to the Gibbs-Boltzmann weight
$\exp(-\beta H)$. This then gives
\begin{equation}
\langle \nabla_{y_i}\phi({\bf y}){\delta H\over \delta\phi({\bf x})}\rangle\vert_{{\bf x}={\bf y}}
= {1\over \beta}\nabla_i\delta({\bf 0})=0
\end{equation}
where the last step can be justified by thinking of the Dirac delta function as the limit of a
Gaussian.  We thus recover the equilibrium result
\begin{equation}
\langle F_i\rangle = \langle \int_{V_S} d{\bf x}\ \nabla_j T_{ij}\rangle =\langle \int_S T_{ij}\ dS_j\rangle.
\end{equation}  

\subsection{Dynamics}
We now consider the dynamical problem where the system in prepared in a state $\phi=0$ at
the time $t=0$ (this could have been by cooling the system to a very low temperature for instance)
and then letting it relax at some non-zero temperature $T$. We will consider case of a general relaxational dynamics, where the evolution of the field is given by
\begin{eqnarray}
{\partial \phi({\bf x})\over \partial t} &=& -\int d{\bf x}'\ R({\bf x},{\bf x}'){\delta H\over \delta\phi({\bf x}') } + \eta({\bf x},t) \nonumber \\
&=&- \int d{\bf x}'\ R\Delta({\bf x},{\bf x}') \phi({\bf x}') +\eta({\bf x},t)  \label{dynamic}
\end{eqnarray}
where $R\Delta$ indicates the composed operator
$R\Delta({\bf x}, {\bf x}')= \int d{\bf y}R({\bf x},{\bf y})\Delta({\bf y},{\bf x}').$ To satisfy detailed balance with 
noise that is uncorrelated in time we chose the noise correlation to be
\begin{equation}
\langle \eta({\bf x},t )\eta({\bf x}',t')\rangle = 2T\delta(t-t')R({\bf x}, {\bf x}'). 
\end{equation}

For the case where $R({\bf x},{\bf x'})=\delta({\bf x}-{\bf x}')$ we recover the case of nonconserved 
model A dynamics and when $R({\bf x},{\bf x'})=-\nabla^2\delta({\bf x}-{\bf x}')$ we have the case
of conserved model B dynamics. In what follows the calculation is valid for any self-adjoint operator
$R$.

The formal solution to this equation, for flat initial configuration of the field,   is
\begin{equation}
\phi({\bf x},t) = \int_0^t ds d{\bf y} \exp\left(-(t-s)R\Delta\right)({\bf x},{\bf y})\eta({\bf y},s)
\end{equation}
This means that the equal time correlation function of the field is given by, in explicit non-operator notation,
\begin{eqnarray}
&&\langle \phi({\bf x},t)\phi({\bf x}',t)\rangle =C({\bf x}, {\bf x}',t) \nonumber \\
&=& 2T \int_0^t  ds d{\bf y}d{\bf y}'
\exp\left(-(t-s)R\Delta\right)({\bf x},{\bf y})\exp\left(-(t-s)R\Delta\right)({\bf x}',{\bf y}')R({\bf y},{\bf y}')  
\end{eqnarray}

Now we use the fact that the operators $\Delta$ and $R$ are self adjoint to write
\begin{equation}
\exp\left(-(t-s)R\Delta\right)({\bf x}'.{\bf y}')= \exp\left(-(t-s)\Delta R\right)({\bf y}',{\bf x}')
\end{equation}

This thus enables us to write in operator notation that
\begin{equation}
C(t)= 2T\int_0^t  ds \exp\left(-(t-s)R\Delta\right) R \exp\left(-(t-s)\Delta R\right)
\end{equation}
Now if we expand the exponential operators in the integral we find
\begin{eqnarray}
\exp\left(-(t-s)R\Delta\right) R \exp\left(-(t-s)\Delta R\right)&=& \sum_{n,m}{1\over n!m!}(-(t-s))^{m+n}(R\Delta)^n R
(\Delta R)^m \nonumber \\ 
&=& \sum_{n,m}{1\over n!m!}(-(t-s))^{m+n}R(\Delta R)^n (\Delta R)^m 
\nonumber \\ &=&R\exp\left(-(t-s)2\Delta R\right)
\end{eqnarray}
The time integration can now be carried out to yield
\begin{equation}
C(t) = T\Delta^{-1}\left[ 1- \exp(-2t\Delta R)\right]
\end{equation}

Now if we Laplace transform this equation (defining $ {\cal L} f(s) = \int_0^\infty dt \exp(-st) f(t)$)we find that
\begin{equation}
{\cal L }C(s) = {T\over s}\left[ \Delta  +{sR^{-1}\over 2}\right]^{-1}\label{ltC}
\end{equation}

Now using this and Eq. (\ref{eqforce}) we find that the Laplace transform for the average value generalized force is given by
\begin{equation} 
\langle {\cal L} F_l(s)\rangle= -{T\over 2s} \int d{\bf x}d{\bf x}' \ \left[{\partial \over \partial l} \Delta({\bf x}, {\bf x}', l)\right] \left[ \Delta  +{sR^{-1}\over 2}\right]^{-1}({\bf x},{\bf x'})
\end{equation}
However as $R$ does not depend on $l$ we may write
\begin{equation} 
\langle {\cal L} F_l(s)\rangle= -{T\over 2s} \int d{\bf x}d{\bf x}' \ \left[{\partial \over \partial l} (\Delta  +{s R^{-1}\over 2})\right]  ({\bf x}, {\bf x'})\left[ \Delta  +{sR^{-1}\over 2}\right]^{-1}({\bf x},{\bf x'})
\end{equation}

Now using the equivalence between  using Eq, (\ref{eqfe}) and Eq. ({\ref{eqfc}) we may write
\begin{equation}
\langle {\cal L} F_l(s)\rangle  =  {T\over s}{\partial\over \partial l}\ln(Z(\Delta_s)) ,\label{main}
\end{equation}
where the operator $\Delta_s$ is given by
\begin{equation}
\Delta_s= \Delta + {s\over 2}R^{-1}
\end{equation}
This result is quite remarkable - it means that the Laplace transform of the time dependent Casimir force
considered here is given by a static Casimir force for another free field theory. It is clear that the result is also valid for the force  on any surface in the system
which interacts with the field. Providing the static partition function is known for the  corresponding static problem, the corresponding time dependent force can be extracted by inverting the Laplace transform.
If one takes the limit $s\to 0$ in Eq. (\ref{main}), we find that the static result is recovered from the pole
at $s=0$ \cite{ben}.

\section{Model A Dynamics}

In this section we will analyze the case of model A dynamics {\em i.e.} where the dynamical operator
$R({\bf x}-{\bf x}') = \delta({\bf x}-{\bf x}')$. This is the easiest case to analyze and it is the case that has been most studied in the literature via the other approaches mentioned in the Introduction.

\subsection{Parallel Plate Geometries}

\begin{figure}
\begin{center}
\resizebox{10 cm}{!}{\includegraphics{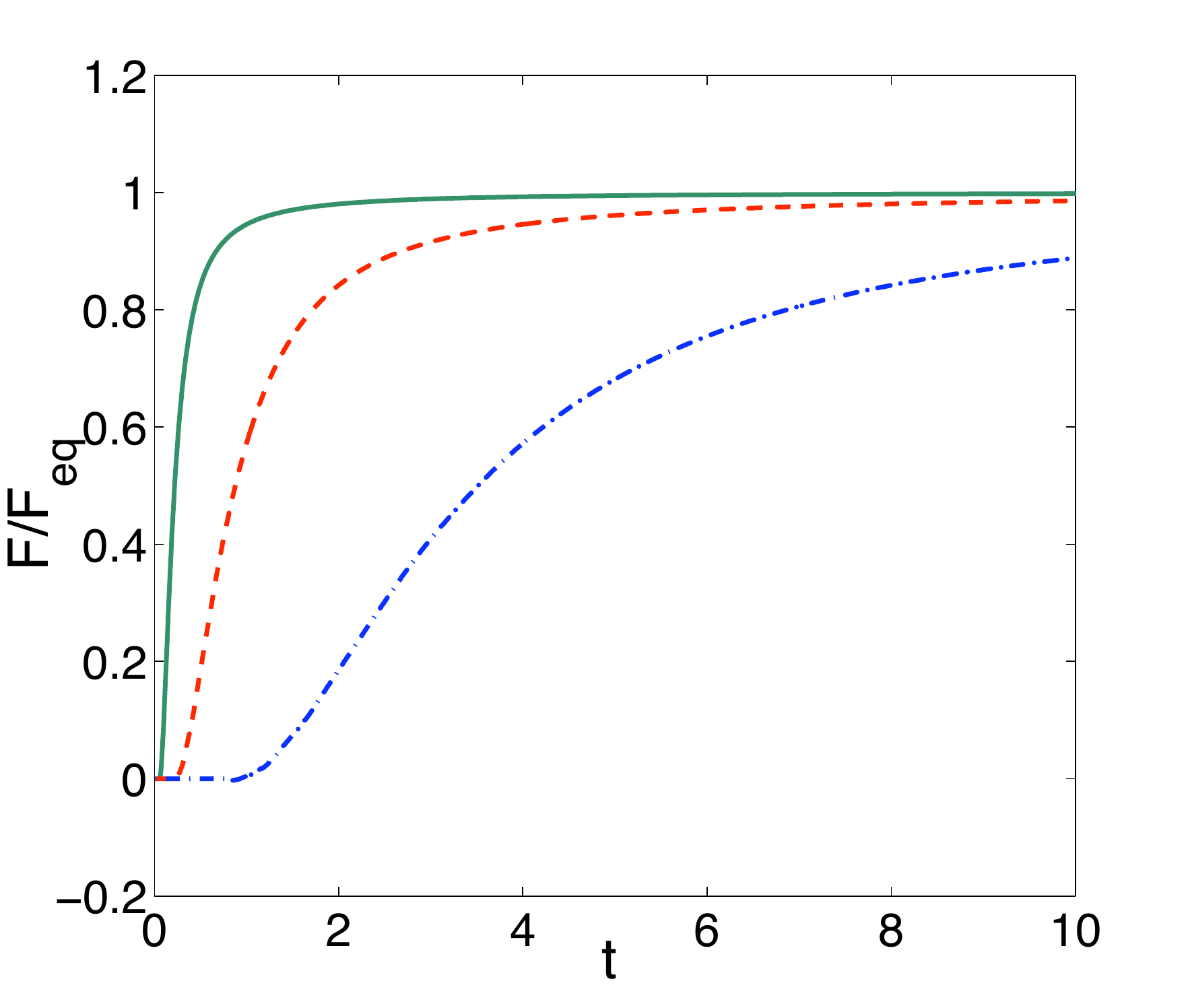}}
\end{center}
\vspace*{-.6cm}
\caption{ For Dirichlet-Dirichlet boundary conditions the approach to equilibrium for various values of the plate separation $l$ (top to bottom $l=1,2,4$) for $d=3$. Nonequilibrium Casimir force $F$, obtained by the direct numerical inverse Laplace transform of Eq.~(\ref{laplaceF}), plotted in units of the equilibrium Casimir force ($F_{eq}$) as a function of time.}
\vspace*{-.6cm}
\label{fig1}
\end{figure}

We now first turn to  the case where the imposed boundary conditions are Dirichlet (DD) and the two plates are immersed in the fluctuating medium (hence there is fluctuating medium on both sides of each plate). 
We take the total length of the system to be $L$ which is fixed, and place the plates of area $A$ 
a distance $l$ apart.   
Standard results on the screened Casimir interaction \cite{kre1992,mos1997,mil2001} give
\begin{equation}
\langle {\cal L} F_l(s)\rangle = -{2AT\over s (4\pi)^{d-1\over 2} \Gamma({d-1\over2})} I(l,s), 
\label{laplaceF}
\end{equation}
where $d$ is the dimension of the space, $A$ is the area of the plates and,
 \begin{equation}
 I(l,s) =  \int k^{d-2} dk  {\sqrt{k^2 + {s\over 2}}\exp(-2l\sqrt{k^2 + {s\over 2}})\over 1-\exp(-2l\sqrt{k^2 + {s\over 2}})},
\end{equation}
The equilibrium behavior is easily extracted by examining the pole at $s=0$ which yields, as anticipated, the standard equilibrium Casimir force
\begin{equation}
\langle F_l\rangle_{eq} =-{AT\Gamma(d)\zeta(d)\over (16\pi)^{{d-1}\over 2}
  \Gamma({d-1\over 2}) l^d},
\end{equation}
where $\Gamma$ is Euler's gamma function and $\zeta$ is the Riemann zeta function \cite{gra2000}
\begin{equation}
\zeta(d)= \sum_{n=1}^\infty {1\over n^d}.
\end{equation}
The full time dependence of the force, starting at zero at $t=0$ and relaxing to the equilibrium value above, can be extracted by direct Laplace inversion of Eq. (\ref{laplaceF}). Fig.(\ref{fig1}) shows the approach to equilibrium for three different plate separations. Clearly, the relaxation times increase with plate separation and this is due to the fact that the underlying dynamics is diffusive and hence $l^2$ sets a time scale. This is also clearly evident in Fig. (\ref{fig2}) that shows the collapsed rescaled force curves obtained by plotting with the time units rescaled by $l^2$.

Useful {\it analytic} expressions for the early and late time behavior of the non-equilibrium force can also be obtained from Eq.(\ref{laplaceF}), because of the technical nature of their derivation they are 
relegated to the Appendix. We find that the temporal derivative of the out of equilibrium force is given by
\begin{equation}
\langle {dF_l\over dt} \rangle 
= -{2AT\over (8\pi)^{d\over 2} t^{d-1\over 2}}{\partial \over \partial t}\sum_{n=1}^\infty 
{1\over \sqrt{t}}\exp(-{l^2 n^2\over 2 t}).\label{nonpo}
\end{equation}
The above expression may also be written in the form
\begin{equation}
\langle {dF_l\over dt} \rangle = -{AT\over 2(8\pi)^{d\over 2} t^{d+2\over 2}}
+ {AT\over 2 (8\pi)^{d-1\over 2}t^{d+1\over 2}}{\partial \over \partial l}\left[\sum_{n=1}^{\infty}\exp(-{2\pi^2n^2 t\over l^2})\right]. \label{poisson}
\end{equation}
Clearly because the underlying dynamics is diffusive $l^2$ sets a time scale.
The short term behavior of the force, $t/l^2 \ll 1$, can be obtained directly from Eq. (\ref{nonpo}) and is given by
\begin{equation}
\langle F_l(t)\rangle  \sim -{2AT\over (8\pi t )^{d\over 2} }\exp(-{l^2\over 2t}).\label{ddst}
\end{equation}
The long time asymptotics, $t/l^2\gg1$,  follow directly from Eq. (\ref{poisson}) and are given by 
\begin{equation}
\langle {F_l} \rangle \sim \langle F_l\rangle_{eq} + {AT\over d(8\pi)^{d\over 2} t^{d\over 2}}.\label{ddlt}
\end{equation}
An interesting thing about Eq. (\ref{poisson}) is that one sees explicitly the appearance of the 
eigenvalues for Dirichlet boundary conditions in the sum on the right hand side. The first term
can be seen to be due to the bulk on the exterior of the system. This can shown  be  by  taking the limit $\l\to \infty$ and expressing the sum as a Riemann integral gives the left hand side to be equal to zero, {\em i.e} as one would expect there is no force. Of course for a system where $l\to\infty$ the system is always out of equilibrium.

The agreement between our asymptotic expressions and exact results obtained by numerical inversion of Eq. (\ref{laplaceF}) are shown in Fig. (\ref{fig2}). As mentioned above the late time correction is independent of $l$. This is because the medium between the two plates has a relaxation time $\tau(l) \sim l^2/2\pi^2$ whereas the slowest relaxation times
in the system are associated with the medium outside the two plates and hence at late times the correction is dominated by the relaxation of the external system in the thermodynamic limit $L\to \infty$. This diffusive relaxation is responsible for the  power law approach to equilibrium. The above result is
also valid for Neumann-Neumann (NN) boundary conditions as the static screened problems have the same force for both DD and NN boundary conditions. We note that this
problem has been studied in \cite{ga2008} {\em assuming} that the  stress tensor can be used to
compute the force out of equilibrium. Our results for DD boundary conditions agree with that given by the stress tensor, though no explicit formula for the force is given in \cite{ga2008} we are able to show
that  this should give the same result for the force (see later). Indeed the numerical curve given for
the DD case in \cite{ga2008} closely resembles ours in Fig. (\ref{fig1}) and the asymptotic form of the late time decay to the equilibrium force agrees with ours.  It is also clear that within this energetic formulation
that the force on the two planes, for any boundary conditions, is equal and opposite as it is given 
via the  {\em equilibrium} force of another free field theory. In \cite{ga2008} the difference between 
the DD and NN boundary conditions can be seen to be due to the presence, for NN
boundary conditions, of the zero (constant) mode in the $z$ direction (perpendicular to the plates) in the computation using the stress tensor.  This is the only difference between  the two results, 
however it is not clear to us how the zero mode can influence the force as it does not see ({\em i.e.} it is not effected by) the NN boundary conditions. 
\begin{figure}
\begin{center}
\resizebox{12 cm}{!}{\includegraphics{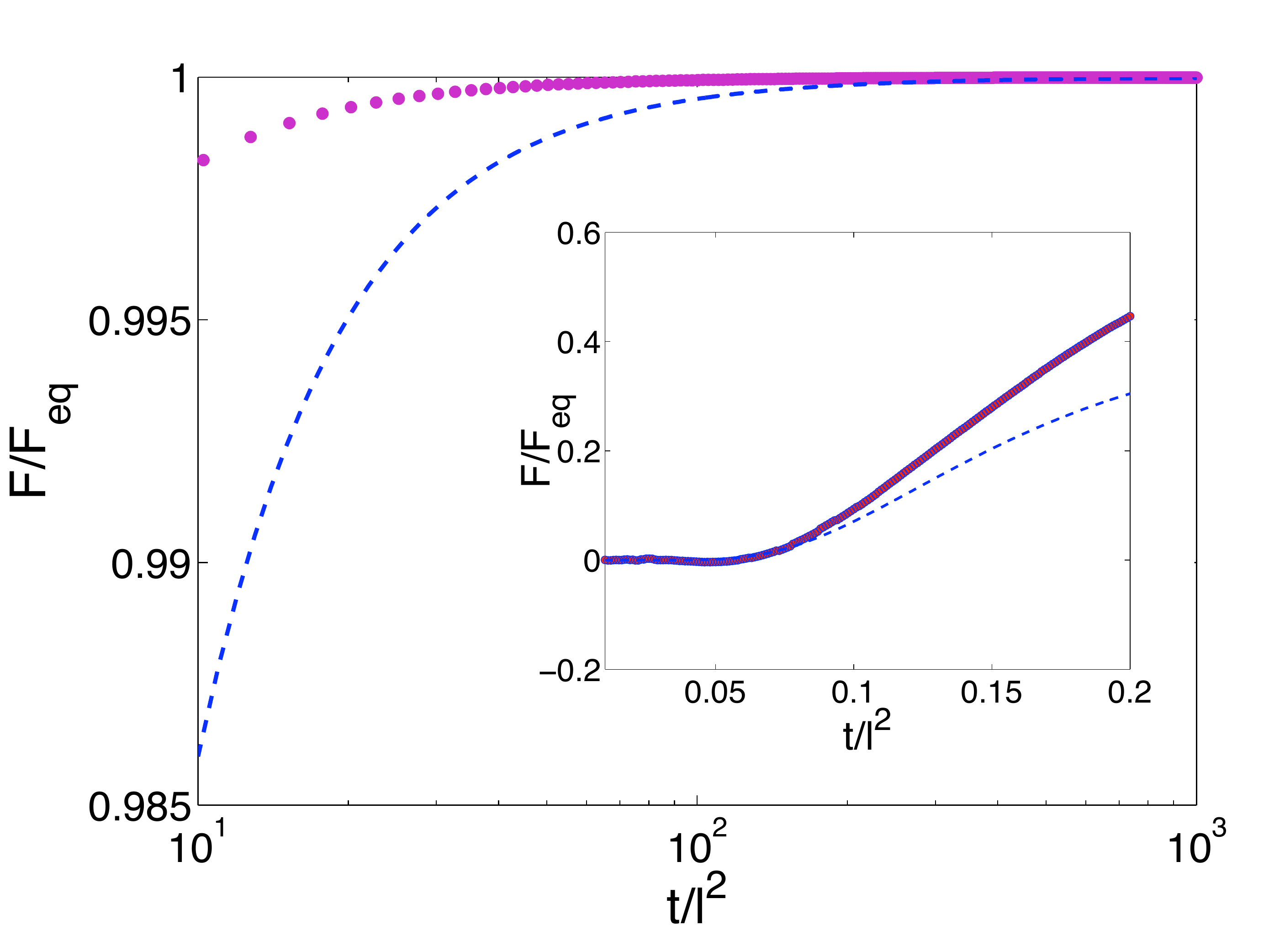}}
\end{center}
\vspace*{-.6cm}
\caption{  Nonequilibrium Casimir force, $F$, in units of the equilibrium Casimir force ($F_{eq}$) at late times and early times (inset) for $d=3$. Symbols were obtained by numerical inverse Laplace transform of Eq.(\ref{laplaceF}). Each plot has only one set of symbols because curves for different $l$ values collapse. Dashed lines are the corresponding approximations from (Eqs.~(\ref{ddst}) and (\ref{ddlt})).}
\vspace*{-.6cm}
\label{fig2}
\end{figure}

To see whether, in general,  a stress tensor computation agrees with our result, we can exploit the
fact that  the right hand side of Eq. (\ref{main}) is for an equilibrium force, for the Hamiltonian 
\begin{equation}
H_{st} = \int d{\bf x}  \ \mathcal{H}_{st},\label{eqHs}
\end{equation}
the subscript $st$ is to make clear that the static or equilibrium measure with Hamiltonian (\ref{eqHs}) is
used to compute observables. Now away from (but still possibly  infinitesimally close to) the boundary the energy density is
given by
\begin{equation}
{\cal H}_{st} = {1\over 2}\left[\nabla\phi\right]^2 + {s\over 4} \phi^2.
\end{equation}  
The stress tensor for this theory is given by
\begin{eqnarray} 
\mathcal{T}_{ij}(s)&=& \delta_{ij} \mathcal{H}_{st} - \nabla_j\phi{\partial\mathcal{H}_{st}\over \partial \nabla_i\phi} \\
&=& {\delta_{ij}\over 2}\left(\left[\nabla\phi\right]^2 + {s\over 2} \phi^2\right) - \nabla_i\phi \nabla_j\phi,
\end{eqnarray} 
and the average force on any volume $V$ is given by the average of integral over the bounding surface $S$ of this
stress tensor
\begin{equation}
\langle F_i(s)\rangle_{st} = \langle \int_S {\cal T}_{ij}(s) dS_j\rangle_{st}
\end{equation}
The Laplace transform of the force on the plates is thus given from Eq.(\ref{main}) by
\begin{equation}
\langle {\cal L} F_l(s)\rangle_{dy} = {1\over s} \langle \int_S {\cal T}_{ij}(s) dS_j\rangle_{st}
\end{equation}
where, to avoid possible confusion the subscript $dy$ indicates averaging over the noise for the dynamical problem and $S$ indicates the surface of the plates. However we can now use the relation Eq. (\ref{ltC}) to write
\begin{equation}
\mathcal{L}( \langle \phi({\bf x},t)\phi({\bf x}',t)\rangle_{dy})(s)  = {1\over s} \langle \phi({\bf x})\phi({\bf x}')\rangle_{st},
\end{equation}
and we may thus write
\begin{equation}
\langle \mathcal{T}_{ij}(s)\rangle_{st} = 
{\delta_{ij}\over 2}\left(\mathcal{L}(\langle \left[\nabla\phi\right]^2\rangle_{dy})(s) + {s\over 2} \mathcal{L}(\langle\phi^2\rangle_{dy})(s) \right) - \mathcal{L}(\langle\nabla_i\phi \nabla_j\phi\rangle_{dy})(s).
\end{equation}
Now we use the fact that if $ \phi({\bf x},0)= 0$ then $\langle \phi({\bf x},0)\phi({\bf x}',0)\rangle_{dy} =0$
and thus 
\begin{equation}
{s\over 2} \mathcal{L}(\langle\phi^2\rangle_{dy})(s)  = {1\over 2}  \mathcal{L}(\langle{\partial
\over \partial t}\phi^2\rangle_{dy})(s) 
\end{equation}
The inverse Laplace transform of the static stress tensor ${\cal T}_{ij}(s)$ can thus be treated as
an effective dynamical stress tensor
\begin{equation}
\langle T^{dy}_{ij}(t)\rangle_{dy} = \mathcal{L}^{-1} \langle \mathcal{T}_{ij}(s)\rangle_{st}(t)
\end{equation}
with
\begin{equation}
T^{dy}_{ij}(t)= {\delta_{ij}\over 2}\left( [\nabla\phi]^2 + {1\over 2} {\partial \phi^2\over \partial t}\right)
-\nabla_i\phi   \nabla_j\phi \label{st}
\end{equation}
and $\phi$ the dynamical field. Therefore computing forces with this dynamical stress tensor will
give the same forces as those given via our boundary energy derivation. Let us emphasize here that
the effective dynamical stress tensor written here is by no means a universal one, {\em it depends
on the precise dynamics of the system} and the {\em choice of initial conditions}.

We see immediately
from Eq. (\ref{st}) that at late time the average value of the time derivative term will go to zero
and we will recover the standard form of the stress tensor . We also see that when computing the force on a plate with Dirichlet boundary conditions the temporal derivative  does not contribute as $\phi$ is zero at the  surface. This is why the results of \cite{ga2008} agree with ours for DD boundary conditions. 

In the case of NN boundary conditions it can be shown that the temporal derivative term cancels out the 
contribution from the zero mode and thus gives exactly the same force as for DD boundary conditions.
Similarly for DN boundary conditions this temporal derivative term ensures that the force at the D boundary is equal in magnitude but opposite to that at the N boundary, this is in contradiction with the
corresponding result given in \cite{ga2008} where the forces at the boundaries are not equal and opposite.

In the case of DN boundary conditions the screened static result is
\begin{equation}
\langle {\cal L} F_l(s)\rangle = -{2AT\over s (4\pi)^{d-1\over 2} \Gamma({d-1\over2})} I(l,s), 
\label{laplaceFDN}
\end{equation}
but here,
 \begin{equation}
 I(l,s) =  -\int k^{d-2} dk  {\sqrt{k^2 + {s\over 2}}\exp(-2l\sqrt{k^2 + {s\over 2}})\over 1+\exp(-2l\sqrt{k^2 + {s\over 2}})},
\end{equation}
 The equilibrium behavior is easily extracted by examining the pole at $s=0$ which yields, as anticipated, the standard equilibrium Casimir force for DN boundary conditions
\begin{equation}
\langle F_l\rangle_{eq} ={AT\Gamma(d)\zeta^*(d)\over (16\pi)^{{d-1}\over 2}
  \Gamma({d-1\over 2}) l^d},
\end{equation}
where \cite{gra2000}
\begin{equation}
\zeta^*(d) = \sum_{n=1}^\infty {(-1)^{n-1}\over n^d} = (1-2^{1-d})\zeta(d) 
\end{equation}
is positive for $d>1$ and so the force in this case is repulsive. Exactly the same analysis as that given
above can be applied to give
\begin{equation}
\langle {dF_l\over dt} \rangle = {2AT\over (8\pi)^{d\over 2} t^{d-1\over 2}}{\partial \over \partial t}\sum_{n=1}^\infty 
{1\over \sqrt{t}} (-1)^{n-1}\exp(-{l^2 n^2\over 2 t}).
\end{equation}
From this we obtain immediately the short time asymptotic behavior of the force
\begin{equation}
\langle F_l(t)\rangle  \sim {2AT\over (8\pi t )^{d\over 2} }\exp(-{l^2\over 2t}),
\end{equation}
we see it is thus opposite that of the case of DD and NN boundary conditions. The large time 
asymptotic behavior of the force can also be extracted by using another form of the Poisson 
summation formula, namely
\begin{equation}
1 + 2\sum_{n=1}^\infty  (-1)^{n}\exp(-{l^2 n^2\over 2 t})=  {\sqrt{2\pi t}\over l}\left[
\sum_{n=-\infty}^\infty \exp(-{2\pi^2(n+{1\over 2})^2 t\over l^2})\right].
\end{equation}
which in the late time limit gives
\begin{equation}
\sum_{n=1}^\infty  (-1)^{n-1}\exp(-{l^2 n^2\over 2 t})
\sim {1\over 2}
\end{equation}
and thus 
\begin{equation}
\langle {dF_l\over dt} \rangle = -{AT \over 2 (8\pi)^{d\over 2} t^{d+2\over 2}},
\end{equation} 
and hence
\begin{equation}
\langle {F_l} \rangle \sim \langle F_l\rangle_{eq} + {AT\over d(8\pi)^{d\over 2} t^{d\over 2}}.
\end{equation}

Therefore we see that the  time correction to the equilibrium result for DN boundary conditions is exactly the same as that for the DD and NN case. However this means that the intermediate force must overshoot its equilibrium value. Fig (\ref{DNfig}) shows the full time dependent approach to equilibrium for three different plate separations obtained by direct numerical inversion of Eq.(\ref{laplaceFDN}). The overshoots at intermediate times are clearly visible. This agrees with the result of \cite{ga2008} for the force calculated at the 
Dirichlet wall as it should from our arguments above stating that one can, as was done in \cite{ga2008}, 
use the usual expression for the stress tensor at the plate with Dirichlet boundary conditions.    This overshoot may possibly be explained in the following way, the effect of the external medium for both types of boundary conditions is to cause an additional, temporally decaying, repulsion between the plates, this is irrespective of whether the equilibrium force between the plates is repulsive or attractive. Thus in the case where the equilibrium force is repulsive we will clearly have an overshoot effect.  This picture is backed up by the fact that the decay is independent of the distance between the plates. 
\begin{figure}
\begin{center}
\resizebox{12cm}{!}{\includegraphics{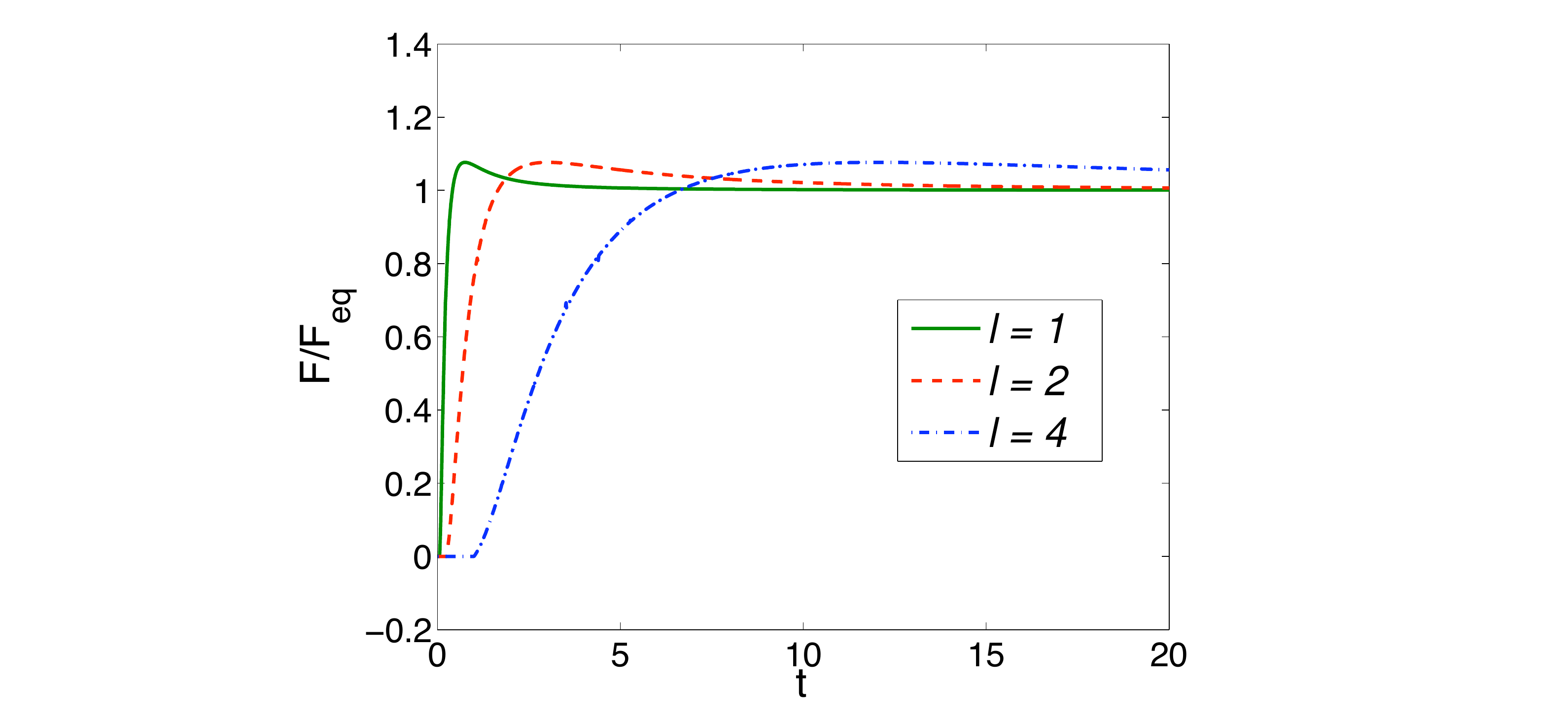}}
\end{center}
\vspace*{-.6cm}
\caption{ For Dirichlet-Neumann boundary conditions, the approach to equilibrium for various values of the plate separation $l$  for $d=3$. Nonequilibrium Casimir force $F$, obtained by the direct numerical inverse Laplace transform of Eq.~(\ref{laplaceFDN}), plotted in units of the equilibrium Casimir force ($F_{eq}$) as a function of time.}
\vspace*{-.6cm}
\label{DNfig}
\end{figure}

\subsection{Effect of Temperature Changes}
The method can also be used to examine the dynamics resulting from a sudden change in temperature, from say $T_0$ where the system is in equilibrium to a temperature $T$. In this case, the initial configuration of the field $\phi({\bf x},0)$ has the correlation function
\begin{equation}
\langle \phi({\bf x},0)\phi({\bf x}',0)\rangle = T_0\Delta^{-1}({\bf x},{\bf x'},l).
\end{equation}
Solving the equation of motion Eq.(\ref{dynamic}) with this initial condition yields the time
dependent correlation function
\begin{equation}
 C({\bf x},{\bf x}',t)= T_0 \exp(-2t \Delta)\Delta^{-1} + T (1-\exp(-2t \Delta))\Delta^{-1}
\end{equation}
and where the second term is exactly the same as that arising for flat initial conditions. Taking the Laplace transform of this gives $s \mathcal{L}C({\bf x},{\bf x}',t)(s) =  {T_0} \left[  \Delta^{-1}({\bf x}, {\bf x},l') -  \Delta^{-1}_s({\bf x}, {\bf x},l')\right] +{T} \Delta^{-1}_s({\bf x}, {\bf x},l')$.
and  we find
\begin{equation}
\langle {\cal L} F_l(s)\rangle =  {T_0 \over s}{\partial\over \partial l}\ln(Z(\Delta)) 
+{T-T_0\over s}{\partial\over \partial l}\ln(Z(\Delta_s)) 
\end{equation}
where we have used the fact that ${\partial\over \partial l}\ln(Z(\Delta_s))$ is independent of the 
temperature. For DD boundary conditions this gives the  limiting behaviors
\begin{eqnarray}
\langle F_l(t)\rangle  &\sim&  \langle F_L\rangle_{eq\ T_0} -{2A(T-T_0)\over (8\pi t )^{d\over 2} }\exp(-{l^2\over 2t}) \ \ {\rm for} \ \ {l^2\over t} \gg 1  \nonumber \\
&\sim& \langle F_L\rangle_{eq\ T} +{A(T-T_0)\over d(8\pi)^{d\over 2} t^{d\over 2}} \ \ {\rm for} \ \ {l^2\over t} \ll 1 
\end{eqnarray}

\section{Non-equilibrium Steady state force - model A forcing with colored noise }
One can also consider the behavior of the Casimir force for relaxational dynamics where the 
deterministic forcing term is of model A type but where the forcing noise is {\it colored} in time such that
$\langle \eta( {\bf x},t)\eta({\bf x}',t)\rangle = T\delta({\bf x}-{\bf x}') \omega\exp(-\omega |t-t|')$ {\em i.e}
\begin{equation}
{\partial \phi\over\partial t} = -{\delta H\over \delta \phi} + \eta
\end{equation}

Here 
$T$ represents an energy scale, $\omega$ a frequency and the resulting steady state is not an equilibrium one. The average 
value of the force in the steady state regime can be computed using the same formalism above and we find that the correlation function of the field is given by
\begin{equation}
C({\bf x},{\bf x'}, \omega ) = T\left[ \Delta({\bf x},{\bf x'}, l)^{-1} - \Delta_{2\omega}({\bf x},{\bf x'}, l)^{-1}\right]
\end{equation}
which yields
\begin{eqnarray}
\langle F_l(\omega)\rangle &=& -{1\over 2}\int d{\bf x} d{\bf x}' {\partial \over \partial l} \Delta({\bf x}, {\bf x}', l) C({\bf x},{\bf x'}, \omega ) \nonumber \\
     &=& {T}{\partial\over \partial l}\left[\ln(Z(\Delta)) -\ln(Z(\Delta_{2\omega})\right] \label{coln}
\end{eqnarray} 

Hence again we find that one can compute a force in a non equilibrium system from  knowledge of static screened systems \cite{comment}. Note that in the limit $\omega \to \infty$ we recover the white noise equilibrium  result of Eq. ({\ref{eqfe}). Fig. (\ref{fig3}) shows the frequency dependence of the non-equilibrium force obtained from Eq.(\ref{coln}) for a two plate system with DD boundary conditions.
This result is also that same as that for NN boundary conditions by the equivalence of the 
corresponding static problems. Again $l^2$ sets a timescale and we see that for $\omega \gg l^{-2}$ the force, $F$ tends to the equilibrium white noise value, $F_{eq}$, as expected, while for $\omega \ll l^{-2}$, $F \ll F_{eq}$ and as $\omega \to 0$, $F $ vanishes. The inset to Fig. (\ref{fig3}) shows how the force depends on plate separation for fixed $\omega$. Again, equilibrium behavior is recovered for large $l$ ($\omega \gg l^{-2}$), while for small plate separations the force changes qualitatively scaling as $l^{-1}$. We note that the result Eq.(\ref{coln}) agrees with a computation for the same system where the steady state Casimir force was computed using the stress tensor \cite{ba2003}. Fig.(\ref{fig4}) shows the frequency dependence of the non-equilibrium force obtained from Eq.(\ref{coln}) for a two plate system with DN boundary conditions. One can see that the qualitative behavior is the same as for DD boundary conditions, tending to zero and the equilibrium white noise value for small and large $\omega$ respectively. In contrast, the scaling of $F$ for small $\omega$ is different. This is most clearly manifested if we look at how the force depends on the plate separation at fixed $\omega$ as shown in the inset to Fig.(\ref{fig4}). While equilibrium behavior is recovered for large $l$ ($\omega \gg l^{-2}$), for small plate separations the force changes qualitatively becoming almost insensitive to the plate separation. This difference is again highlighted in Fig.(\ref{fig5}) which shows an explicit comparison between the curves in Fig. (\ref{fig3}) and Fig. (\ref{fig4}). 
\begin{figure}
\begin{center}
\resizebox{12 cm}{!}{\includegraphics{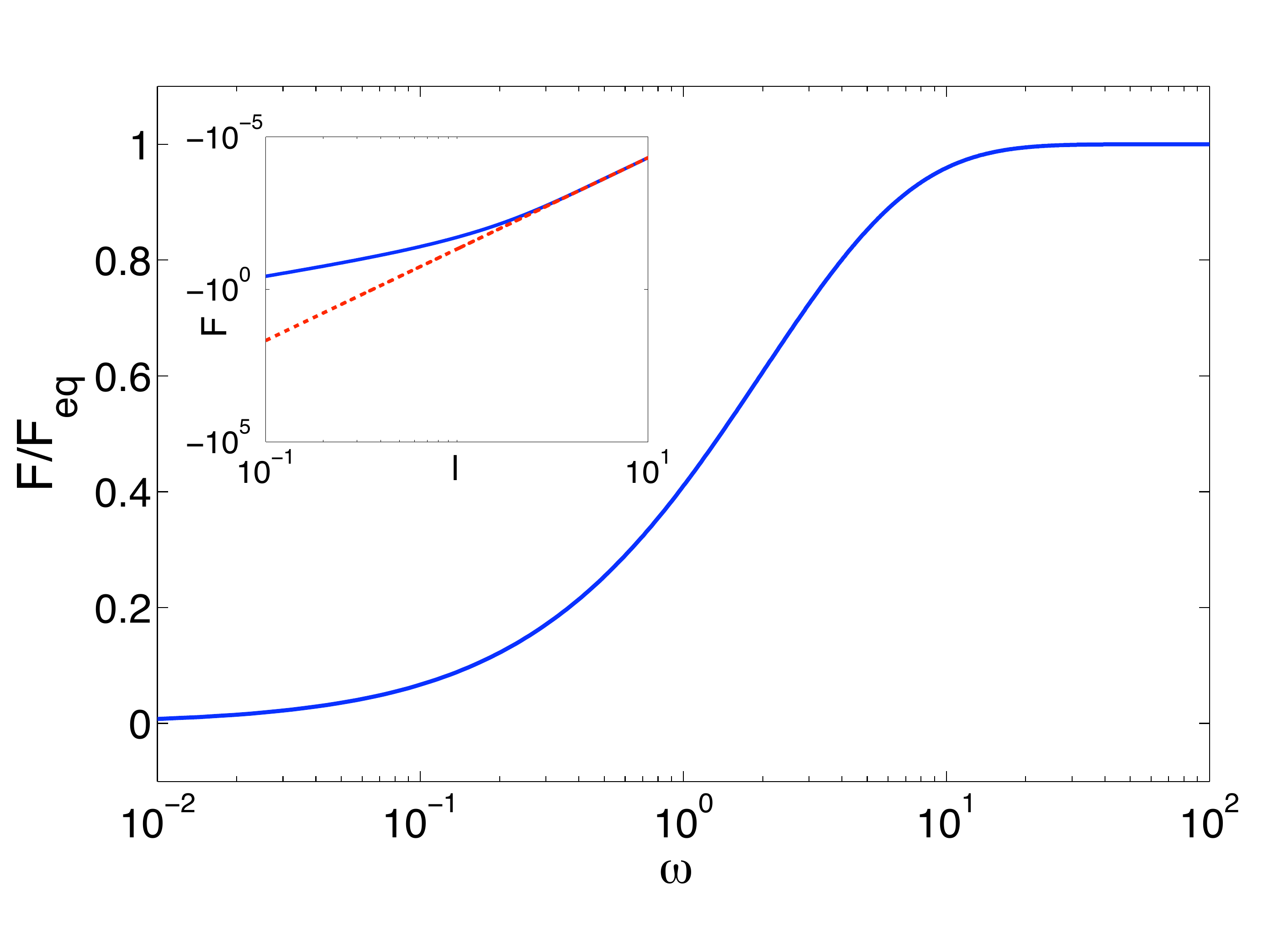}}
\end{center}
\vspace*{-.6cm}
\caption{ Steady state pseudo-Casimir force, $F$, for colored noise with Dirichlet-Dirichlet boundary conditions in units of the equilibrium Casimir force ($F_{eq}$) as a function of $\omega$ (in units of $l^{-2}$) for $l=1$ and $d=3$. Inset: Force per unit area, $F$ (in units of $k_BT/l_0^3$, where $l_0$ is the unit of length),  as a function of plate separation $l$ (in units of $l_0$) for $\omega=1$ ((in units of $l_0^{-2}$)) and $d=3$ (solid line). The equilibrium Casimir force (dashed line) is shown for comparison.}
\label{fig3}
\end{figure}
\begin{figure}
\begin{center}
\resizebox{12 cm}{!}{\includegraphics{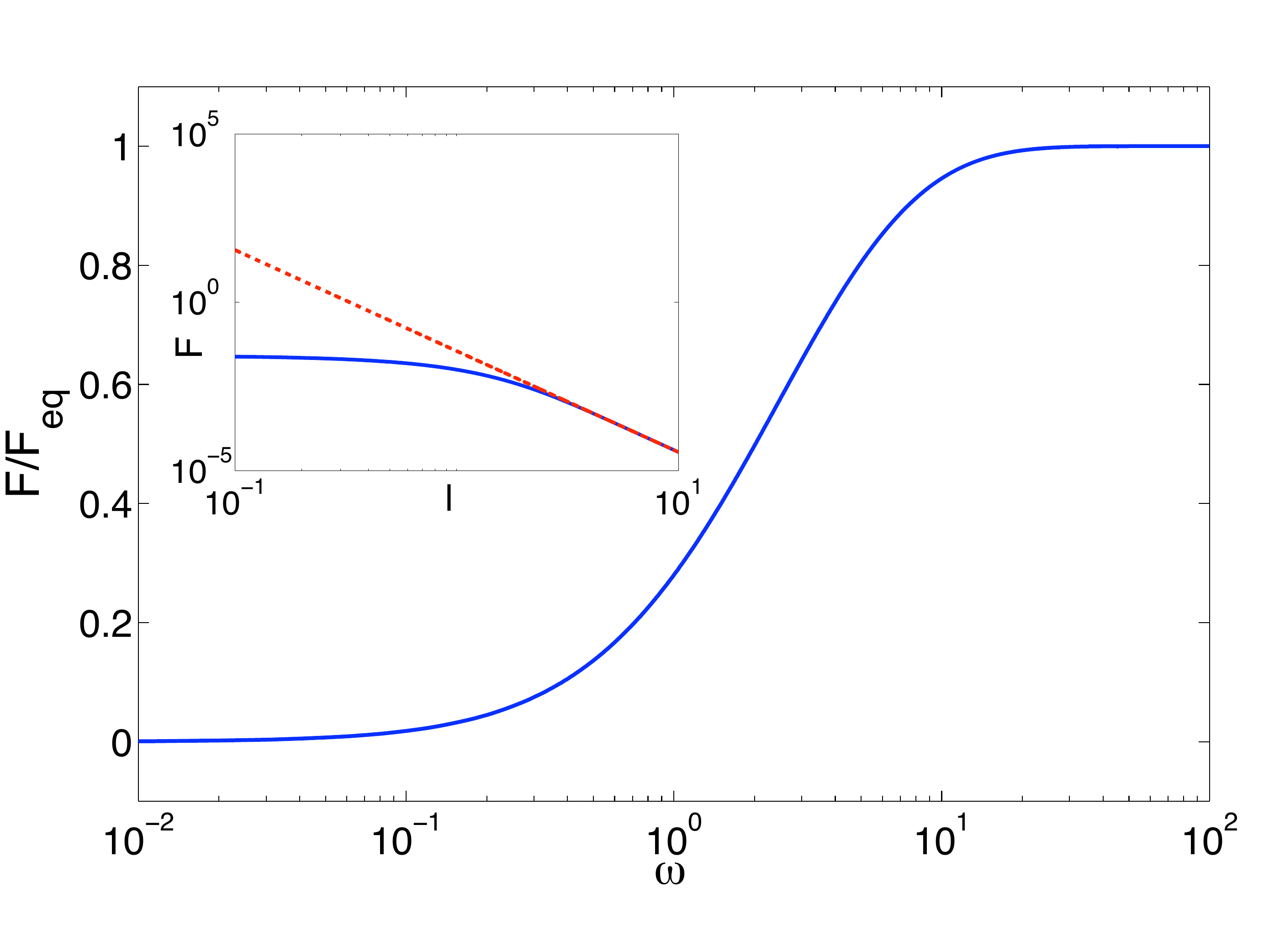}}
\end{center}
\vspace*{-.6cm}
\caption{ Steady state pseudo-Casimir force, $F$, for colored noise with Dirichlet-Neumann boundary conditions in units of the equilibrium Casimir force ($F_{eq}$) as a function of $\omega$ (in units of $l^{-2}$) for $l=1$ and $d=3$. Inset: Force per unit area, $F$ (in units of $k_BT/l_0^3$, where $l_0$ is the unit of length), for Dirichlet-Neumann boundary conditions, as a function of plate separation $l$ (in units of $l_0$) for $\omega=1$ ((in units of $l_0^{-2}$)) and $d=3$ (solid line). The equilibrium Casimir force (dashed line) is shown for comparison.}
\vspace*{-.3cm}
\label{fig4}
\end{figure}
\begin{figure}
\begin{center}
\resizebox{10 cm}{!}{\includegraphics{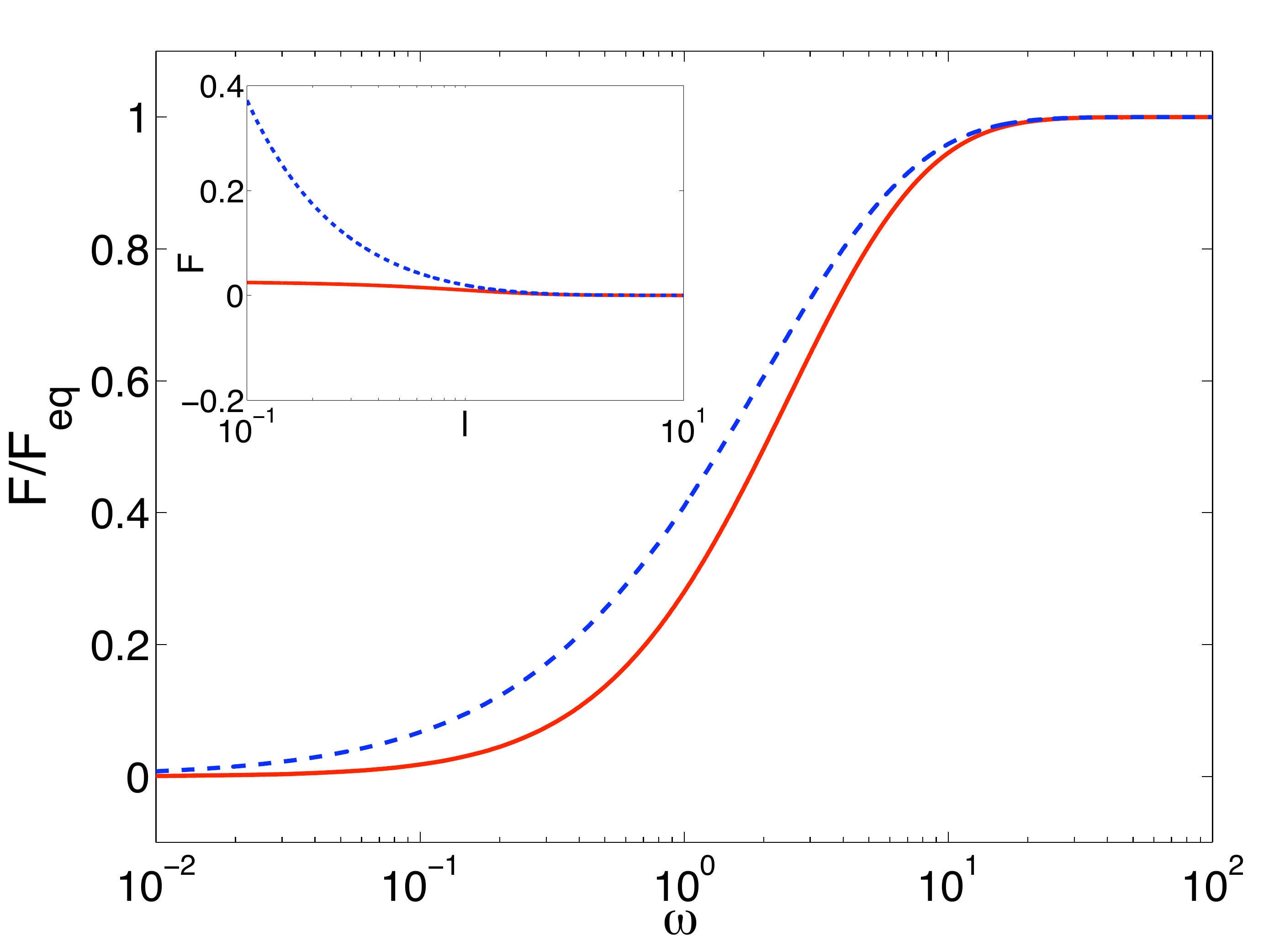}}
\end{center}
\vspace*{-.6cm}
\caption{ Explicit comparison of the steady state pseudo-Casimir force, $F$, for colored noise with Dirichlet-Neumann (red solid line) and Dirichlet-Dirichlet (blue dashed line) boundary conditions in units of the equilibrium Casimir force ($F_{eq}$) as a function of $\omega$ (in units of $l^{-2}$) for $l=1$ and $d=3$ . Inset : Comparison of the absolute value of the force per unit area, $F$ (in units of $k_BT/l_0^3$, where $l_0$ is the unit of length), for Dirichlet-Neumann (red solid line) and Dirichlet-Dirichlet (blue dashed line) boundary conditions, as a function of plate separation $l$ (in units of $l_0$) for $\omega=1$ ((in units of $l_0^{-2}$)) and $d=3$.}
\vspace*{-.6cm}
\label{fig5}
\end{figure}

\section{Pairwise approximation for small defect regions}

We can also study the force between two small {\em defect} regions in an elastic fluctuating medium
using  the pairwise approximation which neglects n-body effects. For instance one could have two small volumes $V_1$ and $V_2$ separated by a distance $l$ in which the field acquires a mass. This will give a Hamiltonian of the form
\begin{equation}
H = {1\over 2}\int d{\bf x}\  [\nabla\phi({\bf x})]^2 + {c_1\over 2}\int_{V_1}  d{\bf x} \ \phi^2({\bf x})
+ {c_2\over 2}\int_{V_2}  d{\bf x} \ \phi^2({\bf x})
\end{equation}
The pairwise approximation is equivalent to evaluating the partition function for the Hamiltonian 
above to second order in the cumulant expansion.  We find that the effective equilibrium potential arising 
between the two regions, when $l$ is much greater than their sizes, is
\begin{equation}
V(l) = -{Tc_1c_2 V_1V_2\over 2} G_0^2(l),\label{eqpw}
\end{equation}
where $G_0 = -\nabla^{-2}$ is the unscreened Coulomb potential in $d$ dimensions. Note that in 
Eq. ({\ref{eqpw}) it is the function $G_0$ evaluated at $l$ which is squared, not the operator . Now we consider how this force evolves towards its static value from the initial conditions where $\phi=0$ throughout the system. Applying the theory developed above we find that, in this same pairwise approximation,  the time dependent force between the two defects is given as the derivative of a time dependent potential $V(l,t)$ whose Laplace transform is given by
\begin{equation}
{\cal L}V(l,s)=   -{Tc_1c_2 V_1V_2\over 2s} G_s^2(l),
\end{equation}
where $G_s =( -\nabla^2 + {s\over 2})^{-1}$. An integral representation can be found 
for arbitrary dimension $d$, however the result in $d=3$ takes the particularly simple form 
\begin{equation}
V(l,t) = -{Tc_1c_2 V_1V_2\over 32\pi^2l^2}{\rm erfc}({l\over \sqrt{2t}}),
\end{equation}
where erfc is the complementary error function.
The effective interaction between the above types of defects for colored driving noise can also
be derived via Eq. ({\ref{coln}). Within the pairwise approximation we find that the steady state force in this case is obtained from the effective potential 
\begin{equation}  
V(l,\omega) = -{Tc_1c_2 V_1V_2\over 2}[G_0^2(l) - G_{2\omega}^2(l)].
\end{equation}
and for $d=3$, for example, we find 
\begin{equation}  
V(l,\omega) = -{Tc_1c_2 V_1V_2\over 32\pi^2l^2}(1-\exp(-2\sqrt{\omega} l)).
\end{equation}
Hence at large separations $l^2\omega \gg1$ the interaction is the same as the that  for white
noise but when $ l^2\omega \ll 1$ the effective potential  behaves as $-\sqrt{\omega}/l$.

\section{Conclusion}
 In this paper we have studied how the Casimir interaction due to a fluctuating scalar field between two
 plates behaves as a function of time in the case of relaxational dynamics, driven by white noise and  obeying detailed balance. In particular we have shown  how it evolves toward its equilibrium value from an initial state where all the field fluctuations are suppressed. We have also analyzed the steady state  
 behavior of the force induced when the field dynamics does not obey detailed balance, notably we have 
 analyzed what happens when the deterministic part of the dynamics is relaxational but the noise
 is colored. 
 Previous studies on the dynamical Casimir effect concentrated on steady state non equilibrium dynamics or dynamics close to equilibrium and considered Dirichlet boundary conditions assuming that the equilibrium stress tensor could be applied to compute the force. Our formalism marks a major advance that overcomes these restrictions by allowing the time-dependent force to be evaluated 
 unambiguously via an expression for the energy of the field. 
The method presented is very general and it would be interesting to analyze the temporal behavior
of the Casimir force for other types of dynamics. We have restricted ourselves to model A type
dynamics as the resulting static results necessary to extract the temporal behavior of the force
are known. The interested reader will see that, for instance, in  the case of model B dynamics
one must know how to compute the Casimir force with a non-local Gaussian action. In addition one must also determine whether the model B dynamics conserves the order parameter within the plates or conserves it globally. Our method applies to the case where the dynamics is globally conserved as
the operator $R$ is assumed to be independent of the plate positions.  The interested reader can
find a discussion of these points in \cite{modelb}.
Another extension of the results here would be to actions with higher order derivative terms such as
the Helfrich action for membrane fluctuations \cite{hel1973}. The main problem here is that the corresponding static results are very complicated even in the case of planar geometries
\cite{deho2007}. However the method should be relatively straightforward to apply in the context
of the pairwise approximation for the interaction between membrane inclusions \cite{nepi1995,netz1997,dema2006} which can be used to analyze the interaction of defect regions of different elasticity and bending rigidity in membranes. In addition the method used here can be applied
to Brownian hydrodynamical dynamics which is the relevant dynamics for membrane fluctuations
\cite{lin2005}.
  
 This research was supported in part by the National Science Foundation under Grant No.PHY05-51164 (while at the KITP UCSB  program
{\em The theory and practice of fluctuation induced interactions} 2008). DSD acknowledges support from   the Institut Universitaire de France. AG acknowledges support from a James S. McDonnell Foundation Award.

\appendix
\section{Analysis of time dependent force for parallel plate geometry with model A dynamics}

Here we carry out an analytical inversion of the Laplace transforms in  Eq. (\ref{laplaceF})
for DD boundary conditions  parallel plate geometry under model A dynamics. We will use a number of standard textbook properties of Laplace transforms which can be found for instance in \cite{ben}.
The computation for ND computations follow with only minor variations.

Starting with the DD case, using the fact that 
$F_l(0)=0$ we may write Eq. (\ref{laplaceF}) as
\begin{equation}
\langle {\cal L}{dF_l\over dt}(s)\rangle = -{2AT\over \sqrt{2}(4\pi)^{d-1\over 2} \Gamma({d-1\over2})}
\int k^{d-2} dk  \sqrt{2k^2 + s}\sum_{n=1}^\infty\exp(-\sqrt{2}ln\sqrt{2k^2 + s}).
\end{equation}
Now we can use the result
\begin{equation}
{\cal L}[{\exp(-pt) f(t)}](s) = {\cal L}f (s+p)
\end{equation}
to obtain
\begin{equation}
\langle {\cal L}{dF_l\over dt}(s)\rangle = -{2AT\over \sqrt{2}(4\pi)^{d-1\over 2} \Gamma({d-1\over2})}
\int k^{d-2} dk  \sum_{n=1}^\infty {\cal L}[{\exp(-2k^2t) f_n(t)](s)}.
\end{equation}
where the Laplace transform of $f_n(t)$ is given by
\begin{equation}
{\cal L}f_n(s) = \sqrt{s}\exp((-\sqrt{2}ln\sqrt{s}).
\end{equation}

We now note that 
\begin{equation}
{\cal L}[{\rm erfc}({a\over \sqrt{ t}})](s) = {1\over s} \exp(-2a\sqrt{s}) \label{eqlt1}
\end{equation}
where erfc denotes the complementary error function defined as
\begin{equation}
{\rm erfc}(z) = {2\over \sqrt{\pi}} \int_z^\infty du\ \exp(-u^2) 
\end{equation}
Eq. (\ref{eqlt1}) can be written as
\begin{equation}
{-1\over 2} {\partial \over \partial a}{\cal L}[{\partial \over \partial t}{\rm erfc}({a\over \sqrt{ t}})](s)
= \sqrt{s}\exp(-2a\sqrt{s}) 
\end{equation}
which thus gives
\begin{equation}
{\cal L}^{-1} \sqrt{s} \exp(-2a\sqrt{s}) = {1\over \sqrt{\pi}}\exp(-{a^2\over t})\left[ -{1\over 2t^{3\over 2}}
+{a^2\over t^{5\over 2}}\right]
\end{equation}
which means that
\begin{equation}
f_n(t) = {1\over \sqrt{\pi}}\exp(-{l^2 n^2\over 2 t})\left[ -{1\over 2t^{3\over 2}}
+{l^2n^2\over  2t^{5\over 2}}\right].
\end{equation}
Putting this together and inverting the Laplace transform yields
\begin{equation}
\langle {dF_l\over dt} \rangle =-{AT\over (8\pi)^{d\over 2} t^{d+2\over 2}}\sum_{n=1}^\infty 
\left[{l^2 n^2\over t} - 1\right]\exp(-{l^2 n^2\over 2 t}),
\end{equation}
which gives Eq. (\ref{nonpo}).
. Short time correspond to the regime where $l^2/t \gg1$ and hence in this regime the dominant behavior is
\begin{equation}
\langle {dF_l\over dt} \rangle= -{ATl^2\over (8\pi)^{d\over 2} t^{d+4\over 2}}\exp(-{l^2\over 2t}).
\end{equation}
Thus at very short times we find that

The long time asymptotics can be obtained by using the Poisson summation formula
\begin{equation}
1 + 2\sum_{n=1}^\infty \exp(-{l^2n^2\over 2t}) = {\sqrt{2\pi t}\over l}\left[
1 + 2\sum_{n=1}^\infty \exp(-{2\pi^2n^2 t\over l^2})\right].
\end{equation}
Using this we find that the time derivative of the average force can be written as Eq. (\ref{poisson})


\begin{thebibliography}{10}
\bibitem{note1}{In writing this we are implicitly assuming that any coarse graining procedure leading to
the effective Hamiltonian is valid at the level of the energy for a given field configuration, it is thus a stronger assumption than that normally made for equilibrium statistical mechanics.}
\end{thebibliography}
\end{document}